\documentclass[a4paper,11pt]{article}
\usepackage{graphicx}
\usepackage{epstopdf}
\usepackage{amsfonts,amsmath,amssymb}
\usepackage{hyperref}

\usepackage{epsfig,multicol,bbm}

\newcommand\fverb{\setbox\pippobox=\hbox\bgroup\verb}
\newcommand\fverbit{\egroup\item[\fbox{\unhbox\pippobox}]}

\newbox\pippobox
\begin{document}
\title{\bf Kaluza-Klein magnetic monopole a la Taub-NUT}
\author{Nematollah Riazi\thanks{Electronic address: n\_riazi@sbu.ac.ir}\,\, and\,\,  S. Sedigheh Hashemi
\\
\small Department of Physics, Shahid Beheshti University, G.C., Evin, Tehran 19839,  Iran}
\maketitle
\begin{abstract}
We present a Taub-NUT Kaluza-Klien vacuum solution and using the standard Kaluza-Klein reduction, show that
this solution  is a static magnetic monopole in 3+1 dimensional spacetime. We find that the four dimensional matter
properties do not obey the equation of state of radiation and
there is no event horizon. A comparison with the available
magnetic monopole solutions, the nature of the singularity,  and the issue of vanishing or negative mass are
discussed.
\end{abstract}

\section{Introduction}\label{sec1}
One of the oldest ideas that unify gravity and electromagnetism is
the theory of Kaluza and Klein which extends space-time to five
dimensions \cite{1}. The physical motivation for this unification
is that the vacuum solutions of the $(4+1)$ Kaluza-Klein field
equations reduce to the $(3+1)$ Einstein field equations with
effective matter and the curvature in $(4+1)$ space induces matter
in $(3+1)$ dimensional space-time \cite{16}. With this idea, the
four dimensional energy-momentum tensor is derived from the
geometry of an exact five dimensional vacuum solution, and the
properties of matter such as density and pressure  as well as
electromagnetic properties are determined by such a solution. In
other words, the field equations of both electromagnetism and
gravity can be obtained from the pure five-dimensional geometry.

Kaluza\rq{}s idea was that the universe has four spatial
dimensions, and the extra dimension is compactified  to form a
circle so small as to be unobservable. Klein's contribution was to
make a reasonable physical basis for the  compactification  of the
fifth dimension \cite{13}. This school of thinking later led to
the eleven-dimensional supergravity theories in  1980s and to the
\lq\lq{}theory of everything\rq\rq{} or ten-dimensional
superstrings \cite{14}.

Many spherically symmetric solutions of Kaluza-Klein type  are investigated in \cite{2}-\cite{7}. In the work  by Gross
and Perry \cite{5} and  Davidson and Owen \cite{6} some other
solutions of the Kaluza-Klein equations are discussed. The $(4+1)$
analogues of $(3+1)$ Schwarzschild  solution are among these
solutions. There are also  solutions of Kaluza-Klein  equations which  do
not have event horizon of  the type which  exists in the
Einstein's theory.

In this paper, we present a  vacuum solution of Kaluza-Klein
theory in five-dimensional spacetime which is closely related to
the Taub-NUT metric. The Taub-NUT  solution has many interesting
features; it carries a new type of charge (NUT charge) which has
topological origins and can be regarded as \lq\lq{}gravitational
magnetic charge\rq\rq{}, so the solution  is known in some other
contexts as the Kaluza-Klein magnetic monopole \cite{ortin}

The plan of this paper is as follows. In section $2$,  we briefly
discuss  the formalism of five dimensional  Kaluza-Klein theory
and the effective four-dimensional Einstein-Maxwell equations. We
will then present a Taub-NUT-like Kaluza-Klien solution and
investigate its physical properties in four dimensions  in section $3$. In the
last section we will draw our main conclusions.

 \section{Kaluza-Klein theory }\label{sec2}
Kaluza $(1921)$ and Klein $(1926)$, used one extra dimension to
unify gravity and electromagnetism in a theory which was basically
five-dimensional general relativity \cite{17}. The theoretical
elegance of this idea is revealed by studying the vacuum solutions
of Kaluza-Klein equations and the matter induced in the
four-dimensional spacetime\cite{18}. Thus, we are chiefly
interested in the vacuum five dimensional Einstein equations.
 For any vacuum solution, the  energy-momentum tensor vanishes and thus $\hat G_{AB}=0$
or, equivalently $ \hat R_{AB}=0 $, where $\hat G_{AB} \equiv \hat
R_{AB}- \frac{1}{2}\hat R \hat g_{AB}$ is the Einstein tensor,
$\hat R_{AB}$ and $\hat R = \hat g_{AB} \hat R^{AB}$ are  the
five-dimensional Ricci tensor and scalar, respectively.  $\hat
g_{AB}$ is the metric tensor in  five dimensions. Here, the
indices $A,B,...$ run over $0...4$ \cite{14}.

Generally, one can identify the $\mu \nu$ part  of $\hat g^{AB}$
with $g^{\mu \nu}$, which is the contravariant four dimensional
metric tensor, $A_{\mu}$ with the electromagnetic potential  and
$\phi$ as a scalar field.  The correspondence between the above
components is
\begin{equation}\label{eq1}
\hat g^{AB}= \left(
‎\begin{array}{ccccccc}‎
‎g^{\mu \nu}‎   & -\kappa A^{\mu} &  \\‎
 ‎‎  &‎\\‎
-\kappa A^{\nu}    ~~& \kappa^2 A^{\sigma}A_{\sigma}+\phi^2 &  \\‎
‎\end{array}
‎\right)
\end{equation}
where $\kappa$ is a coupling  constant for  the electromagnetic potential $A^{\mu}$
and the indices $\mu,\nu$ run over $0...3$.

The five dimensional field equations reduce to the four dimensional field equations \cite{8},\cite{Y}
\begin{align}\label{eq2}
G_{\mu \nu}=\frac{\kappa^2 \phi^2}{2}T_{\mu \nu}^{EM}-\frac{1}{\phi}\left[\nabla_{\mu}(\partial _{\nu} \phi)
-g_{\mu \nu}\Box \phi\right],
\end{align}
and
\begin{align}\label{eq3}
\nabla^{\mu}F_{\mu \nu}=-3\frac{\partial ^{\mu}\phi}{\phi}F_{\mu \nu},  \nonumber\\
\Box \phi= \frac{\kappa^2 \phi^3}{4} F_{\mu \nu} F^{\mu \nu},
\end{align}
where $G_{\mu \nu}$ is the Einstein tensor,
$T_{\mu \nu}^{EM}\equiv \frac{1}{4} g_{\mu \nu}F_{\rho \sigma}F^{\rho \sigma}-F_{\mu}^{~\sigma}F_{\nu \sigma}$
is the electromagnetic energy-momentum tensor
 and  the field strength $F_{\mu\nu}=\partial_{\mu} A_{\nu}-\partial_{\nu}A_{\mu}$. Knowing the five dimensional
 metric, therefore, leads to a complete knowledge of the four dimensional geometry, as well as the electromagnetic
 and scalar fields.
\section{ Taub-NUT solution and Kaluza-Klein magnetic monopole}
The Taub-NUT solution was first discovered by Taub (1951), and
subsequently extended by Newman, Tamburino and Unti (1963) as a
generalization of the Schwarzschild spacetime \cite{21}. This
solution is a single, non-radiating  extension of the Taub universe. It is an anisotropic but spatially homogeneous
vacuum solution  of
Einstein\rq{}s field equations with topology $R^1 \times S^3$.
The Taub metric is given by
\begin{equation}
{\rm d}s^2=-\frac{1}{V(t)}{\rm d}t^2+4b^2V(t)({\rm d}\psi+\cos \theta {\rm d}\phi)^2+(t^2+b^2)({\rm d}\theta^2+\sin \theta^2{\rm d}\phi^2),
\end{equation}
where $V(t)=-1+ 2(mt+b^2)(t^2+b^2)^{-1}$,  $m$ and $b$ are positive constants, $\psi, \phi, \theta$ are Euler angels with usual ranges \cite{k}.
The Taub-NUT solution is nowadays
being used in the context of higher-dimensional theories of
semi-classical quantum gravity \cite{20}. As an  example, in the
work by Gross and Perry \cite{5} and of Sorkin \cite{4},  soliton
solutions were obtained by embedding the Taub-NUT gravitational
instanton inside the five dimensional Kaluza-Klein
manifold\cite{21}. Solitons are static, localized, and
non-singular solutions of nonlinear field equations which resemble
particles. One  such solution which obeys the Dirac quantization
condition is considered in \cite{5}.

The Kaluza-Klein monopole of Gross and Perry is  described by the
following metric which is a generalization of the self-dual
Euclidean Taub-NUT solution \cite{5}
\begin{align}\label{eq4}
 {\rm d}s^2=- {\rm d}t^2+V( {\rm d}x^5+4m(1-\cos\theta) {\rm d}\phi)^2+\frac{1}{V}( {\rm d}r^2+r^2 {\rm d} \theta^2 +r^2 \sin ^2\theta  {\rm d}\phi^2),
\end{align}
where
\begin{align}
\frac{1}{V}=&1+\frac{4m}{r}.
\end{align}
This solution has a coordinate singularity at $r=0$ which is
called NUT singularity. This can be absent if the coordinate $x^5$
is periodic with period $16\pi m=2\pi R$, where $R$ is the radius
of the fifth dimension. Thus $m=\frac{\sqrt{\pi G}}{2e}$\cite{28}.
The gauge field $A_{\nu}$ is given by $A_{\phi}=4m(1-cos\theta)$,
and the magnetic field is $B=\dfrac{4m\bold{r}}{r^3}$, which is
clearly that of a monopole and has a Dirac string singularity in
the range $r=0 $ to $\infty$. The magnetic charge of this monopole
is $g=\dfrac{m}{\sqrt{\pi G}}$
 which has one unit of Dirac charge. In this model,
 the total magnetic flux is constant.
For this solution, the soliton mass is determined to be
$M^2=\dfrac{m_{p}^2}{16\alpha}$ where $m_{p}$ is the Planck mass.
Gross and Perry showed that the Kaluza-Klein theory can contain
magnetic monopole solitons which would support the unified gauge
theories and allow for searching the physics of unification.

Here, we introduce a metric which is a vacuum five dimensional
solution, having some properties in common with the monopole of
Gross and Perry, while some other properties being different. The
proposed static metric is given by
\begin{align}\label{6}
 {\rm d}s^2_{(5)}= - {\rm d}t^2+w(r)\left( {\rm d}r^2+r^2 {\rm d}\theta^2 +r^2\sin^2 \theta  {\rm d}\phi^2\right)+
 \frac{k}{w(r)}\left( {\rm d}\psi + Q\cos\theta  {\rm d}\phi\right)^2.
\end{align}
Here, the extra coordinate is represented by $\psi$.  $k$ and  $Q$ are constants and $w(r)$ is a function (to be determined) of the radial coordinate $r$. The coordinates take on values within the usual ranges:
$r\geq0$, $0\leq\theta\leq\pi$, $0\leq  \phi\leq 2\pi$ and $0\leq
\psi \leq 2\pi$.

The Ricci scalar associated with the five dimensional metric (\ref{6}) is given by
\begin{equation}\label{7}
R=\frac{1}{2}\frac{1}{w(r)^3r^4}\left[2r^4w(r)w''(r)+4w\rq{}(r)w(r)r^3 -w\rq{}(r)^2 r^4+kQ^2\right].
\end{equation}
Since we are interested in vacuum solution where the Ricci tensor is zero $(R_{AB}=0)$ it is necessary but not sufficient to have a zero Ricci scalar $R=0$ which can be solved for the function $w(r)$ to give
\begin{equation}
w(r)=k_1+\frac{k_2}{r}+\frac{k_3}{r^2},
\end{equation}
where $k_1$, $k_2$ and $k_3$ are constants. By substituting  the function $w(r)$ in the Ricci scalar (\ref{7}) and requiring a zero Ricci scalar, one can obtain the following constraint between the constants
\begin{equation}
kQ^2+4k_1k_3-k_2^2=0,
\end{equation}
or
\begin{equation}
k_3=\frac{1}{4k_1}\left(k_2^2-kQ^2\right).
\end{equation}
Thus, the Ricci scalar in terms of the above constants reduces to
\begin{equation}
R=\frac{1}{2}\frac{r^2(4k_1-1)(kQ^2-k_2^2)}{(kQ^2-k_2^2-k_2r-k_1r^2)^3}.
\end{equation}
In order to have a zero Ricci scalar the two following  possibilities remain
 \begin{equation}\label{12}
R=0 \Longrightarrow \begin{cases}
k_1=\frac{1}{4} \Longrightarrow R_{AB}\neq 0 ,\\ {\rm or} \\ 
k_2=\pm \sqrt {k}Q \Longrightarrow R_{AB}= 0, \quad R_{ABCD}\neq0.
\end{cases}
\end{equation}
 As it is seen from Eq. (\ref{12}) with $k_1=\frac{1}{4}$, the Ricci tensor is not zero and does not give a vacuum solution, thus this case is discarded. However, the second choice $k_2=\pm \sqrt {k}Q$ gives a  Ricci flat solution with a non-zero Riemann tensor in five dimensions which is what we are  interested in. Consequently, the general form of the function $w(r)$
will be  given by
\begin{equation}
w(r)=k_1\pm \frac{\sqrt {k}Q }{r}.
\end{equation}
In what follows, we take the constants $k_1=1$, $Q=1$ and $k=4m^2$. Therefore, metric (\ref{6}) reduces to (we choose the minus sign for the second term in $w(r)$)
\begin{align}\label{14}
 {\rm d}s^2_{(5)}= - {\rm d}t^2+(1-\frac{2m}{r})\left( {\rm d}r^2+r^2 {\rm d}\theta^2 +r^2\sin^2 \theta  {\rm d}\phi^2\right)+
\left(\frac{4m^2}{1-\frac{2m}{r}}\right)\left( {\rm d}\psi + Q\cos\theta  {\rm d}\phi\right)^2.
\end{align}
The Killing
vectors associated with metric (\ref{14})   are given
by
\begin{align}
&K_{0}=(1,0,0,0,0),\quad  K_{1}=(0,0,0,0,1), \quad  K_{2}=(0,0,0,1,0),\nonumber\\
&K_{3}=(0,0, -\sin\phi, -\cot\theta \cos \phi, \csc \theta \cos\phi),\nonumber\\
&K_{4}=(0,0, \cos\phi, -\cot\theta \sin\phi, \csc \theta \sin\phi),
\end{align}
which are  the same as in the Taub-NUT space  discussed in
\cite{25}, where the authors studied spinning particles in the
Taub-NUT space.

  The gauge field, $A_{\mu}$ and the scalar field
$\phi$ deduced from the  metric  (\ref{14}) with the help of (\ref{eq1}) are
\begin{equation}
A_{\phi}=\frac{cos\theta}{\kappa},
 \end{equation}
 and
 \begin{equation}
 \phi^2=\frac{4m^2}{1-\frac{2m}{r}},
 \end{equation}
respectively.
 The only non-vanishing component for the electromagnetic field tensor which is related to the gauge field $A_\mu$ via $F_{\mu \nu}=\partial _\mu A_\nu-\partial _\nu A_\mu$ is
\begin{equation}
F_{\theta \phi}=-F_{\phi \theta}= -\frac{sin \theta}{\kappa},
\end{equation}
which corresponds to a radial magnetic field  $B_{r}=\dfrac{1}{\kappa r^2}$ with a magnetic
charge $Q_{M}=\dfrac{1}{\kappa}$. As the radial coordinate  $r$ goes to  infinity $r\rightarrow \infty$, the scalar field
equals $\phi_{0}^2=4m^2$   so that the
second part of equation (\ref{eq2}) becomes zero, therefore
\begin{equation}
G_{\mu \nu}=\frac{\kappa^2\phi_{0}^2}{2}T^{EM}_{\mu \nu}=8\pi GT_{\mu \nu}, \qquad as\qquad r\rightarrow \infty,
\end{equation}
where we have put the speed of light $c$ equal to $1$. By
comparing this equation with the  Einstein equation, we
obtain $\dfrac{\kappa^2 \phi_{0}^2}{2}=8\pi G$, thus the constant
$\kappa$ equals $\kappa=\dfrac{2}{m}\sqrt{\pi G}$.  Straightforwardly, the total magnetic monopole  charge is given by
\begin{equation}\label{eq11}
Q_{M}=\frac{m}{2}\frac{1}{\sqrt{ \pi G}}.
\end{equation}
The total magnetic flux through any spherical surface centered at
the origin is calculated as \cite{26}
\begin{equation}\label{22}
\Phi_{B}=\int F=\frac{1}{2}\oint  F_{\mu\nu}d\Sigma^{\mu
\nu}=\frac{\pi}{\kappa},
\end{equation}
 where $F_{\mu \nu}$ is the electromagnetic field tensor  and ${\rm d}\Sigma^{\mu \nu}$ is an element of two-dimensional surface area.  It is observed from Eq. (\ref{22}) that  the flux of the magnetic monopole is constant
(i.e. we have a singular magnetic charge).

We now  show that  equation (\ref{eq3}) is satisfied explicitly in
four dimensions. We first note that
\begin{equation}\label{eq13}
\Box \phi= \frac{4m^3}{r^4(1-\frac{2m}{r})^\frac{7}{2}}.
\end{equation}
 The value of the scalar $F_{\mu \nu}F^{\mu \nu}$
is easily  calculated to be
\begin{equation}
F_{\mu\nu}F^{\mu \nu}=\frac{2}{\kappa^2 r^4 \left(1-\frac{2m}{r}\right)^2},
\end{equation}
and therefore
\begin{equation}
\frac{\kappa^2 \phi^3}{4}F_{\mu\nu}F^{\mu \nu}=\frac{4m^3}{r^4(1-\frac{2m}{r})^\frac{7}{2}},
\end{equation}
which together  with (\ref{eq13}) shows that equation (\ref{eq3}) is satisfied.\\
The four dimensional metric deduced  from  equation (\ref{14}) with the use of (\ref{eq1})
leads to the following   asymptotically flat spacetime:
\begin{align}\label{26}
 {\rm d}s^2_{\left(4\right)}=- {\rm d}t^2+\left(1-\frac{2m}{r}\right) {\rm d}r^2+r^2\left(1-\frac{2m}{r}\right)\left[ {\rm d}\theta^2 +sin \theta ^2
 {\rm d}\phi^2\right].
\end{align}
 For this metric,
the Ricci scalar  and the Kretschmann  invariant $K$ are given by
\begin{equation}
R=\frac{6m^2}{r^4\left(1-\frac{2m}{r}\right)^3},
\end{equation}
and
 \begin{equation}
 K=R_{\mu \nu \rho \sigma}R^{\mu \nu \rho \sigma}=\frac{m^2}{\left(r-2m\right)^5}\left[\frac{2}{r-2m}+\frac{2r-3m}{r^2}\right].
 \end{equation}
It is seen that the four dimensional metric (\ref{26}) has two {\bf curvature} singularities at $r=0$ and $r=2m$.
 In order to  achieve more insight into the nature of the singularity at $r=2m$, let us calculate the proper surface area of a $S^2$ hypersurface at constant $t$ and $r$; 
\begin{equation}
A\left(r\right)=\int \sqrt{g^{\left(2\right)}} {\rm d}x^2= \int r^2\left(1-\frac{2m}{r} \right)sin \theta  {\rm d} \theta  {\rm d} \phi=
4 \pi  r^2\left(1-\frac{2m}{r} \right).
\end{equation}
It can be seen that the surface area $A\left(r\right)$ becomes
zero at $r=2m$,
 showing that the space shrinks
  to a point at $r=2m$. This result shows that the $r=2m$
  hypersurface is not an event horizon.
For the case $r>2m$ the signature of the metric is proper $(-,+,+,+)$
but for the range $r<2m$ the signature of
 the metric will be improper
and non-Lorentzian $ (-,-,-,-)$, thus the patch
$r<2m$ should be excluded from the spacetime. From now  on,  the spacetime is
only considered in the range  $r\geq2m$.
 Since the range $0<r<2m$ is omitted from the spacetime, there is now only one curvature singularity at
 $r=2m$. In order to see this more clearly, let us transform the metric (\ref{26})  into the following form,
using the radial coordinate transformation ${\tilde r}^2=r^2\left(
1-\frac{2m}{r}\right)$;
\begin{equation}
 {\rm d}s^2_{(4)}=- {\rm d}t^2+\frac{{\tilde r}^4}{(m^2+{\tilde r}^2)(m+
\sqrt{m^2+{\tilde r}^2})^2} {\rm d}{\tilde r}^2 +{\tilde r}^2 {\rm d}\Omega^2.
\end{equation}
While the  metric (\ref{26}) is singular at  $r=0$ and $r=2m$,
the transformed metric is singular only at ${\tilde r}=0$ which is a
curvature naked singularity.

The properties of the induced matter associated with the above
metric can be gained by the equation (\ref{eq3}).
The components of the energy-momentum tensor can be easily
calculated for the metric (\ref{26}). We find that the effective
source is like a fluid with an anisotropic pressure. At
sufficiently large $r$, the energy-momentum components will tend
to zero. The trace of the energy-momentum tensor does not vanish
generally, which shows that the effective matter field around the
singularity can not be considered as an ultra-relativistic quantum
field in contrast to the Kaluza-Klein solitons described in
\cite{23}. The gravitational mass can be obtained from the
energy-momentum tensor components according to
\begin{equation}
M_{g}\left(r\right) \equiv \int_{2m}^r \left(T^{0} _{~0}-T^{1} _{~1}-T^{2} _{~2}-T^{3} _{~3} \right)\sqrt{g^{(3)}} {\rm d}V_{3},
\end{equation}
where ${g^{(3)}}$ is the determinant of the 3-metric and $dV_{3}$
is a $3D$ volume element. One would expect the mass of the
monopole be independent of the sign of the magnetic charge, since
the electromagnetic energy density depends on the magnetic field
squared and independent of its direction. Using the components of
the $T^{\mu}_{~\nu}$ make the value of  integrand zero so that
\begin {equation}
M_{g}\left(r\right)=0,
\end{equation}
meaning that   the gravitational mass vanishes  everywhere. The conserved Komar mass which  is defined as \cite{27}
\begin{equation}\label{eq29}
M_{g}(r)=-\frac{1}{8\pi}\oint _{S}\nabla ^{\mu}K^{\nu} {\rm d}S_{\mu \nu},
\end{equation}
also leads to
\begin{equation}
M_{g}(r)=0,
\end{equation}
which is in agreement with the gravitational mass.

If in the process of compactification from $(4+1)$ to $(3+1)$ dimensions we use the ansatz
\begin{equation}
\hat G_{AB}= \phi^{\beta}\left(
‎\begin{array}{ccccccc}‎
‎g_{\mu \nu}+\phi A_{\mu}A_{\nu}‎   & \phi A_{\mu} &  \\‎
 ‎‎  &‎\\‎
\phi A_{\nu}    ~~& \phi &  \\‎
‎\end{array}
‎\right),
\end{equation}
then the choice of $\beta$ for which $\phi$ does not appear explicitly is called the Einstein frame.
Using the above equation will lead to the four dimensional metric
\begin{align}\label{37}
 {\rm d}s^2_{\left(4\right)}=\phi^{-\beta}\left[- {\rm d}t^2+\left(1-\frac{2m}{r}\right) {\rm d}r^2+r^2\left(1-\frac{2m}{r}\right) {\rm d}\Omega^2\right],
\end{align}
by choosing $\phi^{-\beta}=(1-\frac{2m}{r})^{-\frac{1}{2}}$, equation (\ref{37}) reduces to
\begin{align}\label{38}
 {\rm d}s^2=-(1-\frac{2m}{r})^{-1/2}  {\rm d}t^2+(1-\frac{2m}{r})^{1/2}( {\rm d}r^2+r^2  {\rm d}\Omega^2).
\end{align}
This is the metric obtained by  Gibbons and Manton in \cite{gib} by replacing $m$ with $-m$.
 The metric (\ref{38}) with $-m$ gives a finite  value for total energy which is equal to $E=\frac{m}{2G}$.

\section{Conclusion}
Inspired by  the Taub-NUT solution, we introduced a Kaluza-Klein
vacuum solution in $(4+1)D$, which described a magnetic monopole in
$(3+1)$D spacetime. The gravitational mass of the solution was
shown to vanish, contrary to the magnetic monopole of Gibbons and Manton\cite{gib}. The fluid supporting the four dimensional
space-time was shown to differ from that of an ultra-relativistic
fluid, in contrast with the work by Wesson and Leon\cite{23}. The
pressure is anisotropic in both works.

The main conclusion of this paper is that the presented Taub-NUT
Klauza-Klein solution contains a singular magnetic monopole with
vanishing gravitational mass. We calculated the magnetic charge
and showed that the total magnetic flux of the monopole through
any spherical surface centered at the origin was constant,
indicating that the monopole is singular and there is no extended
magnetized fluid. The gravitational mass was derived in two
different ways and was shown to vanish using both definitions.  It
was pointed out that the singularity which appears at finite $r$
is neither a horizon, nor a surface of finite, non-vanishing
surface area. In a more appropriate coordinate system, this was
shown to be curvature singularity at ${\tilde r}=0$.\\

{\bf Acknowledgements}\\

The authors acknowledge the support of Shahid Beheshti University Research Council.

\end{document}